\documentclass[a4paper,12pt]{amsart}
\usepackage[hmarginratio={1:1},vmarginratio={1:1},heightrounded,textwidth=455pt]{geometry}
\usepackage{subcaption}
\usepackage[usenames,dvipsnames]{xcolor}
\usepackage{amsfonts}
\usepackage{amsmath}
\usepackage{amsthm}
\usepackage[utf8]{inputenc}
\usepackage{amssymb, colonequals}
\usepackage{hyperref}
\usepackage{tikz}
\usetikzlibrary{arrows}
\usetikzlibrary{patterns}
\usetikzlibrary{arrows}
\usetikzlibrary{patterns}
\usetikzlibrary{calc}
\usepackage[textsize=small]{todonotes}
\usepackage{url}
\usepackage{enumitem}
\usepackage{multirow}
\usepackage{chngcntr}
\counterwithin{figure}{section}
\setlist[itemize]{leftmargin=0.4in}

\newtheorem{innercustomgeneric}{\customgenericname}
\providecommand{\customgenericname}{}
\newcommand{\newcustomtheorem}[2]{%
  \newenvironment{#1}[1]
  {%
   \ifdefined\crefalias\crefalias{innercustomgeneric}{#2}\fi
   \renewcommand\customgenericname{#2}%
   \renewcommand\theinnercustomgeneric{##1}%
   \innercustomgeneric
  }
  {\endinnercustomgeneric}%
  \ifdefined\crefname\crefname{#2}{#2}{#2s}\fi
}

\newcustomtheorem{customthm}{Theorem}
\newcustomtheorem{customlemma}{Lemma}

\let\geq\geqslant
\let\setminus\smallsetminus

\let\rho\varrho

\newcommand{\Oh}[1]{\ensuremath{\mathcal{O}(#1)}}

\makeatletter
\newcommand\ie{i.e\@ifnextchar.{}{.\@}}
\newcommand\etc{etc\@ifnextchar.{}{.\@}}
\newcommand\etal{et~al\@ifnextchar.{}{.\@}}
\def\namedlabel#1#2{\begingroup
    #2%
    \def\@currentlabel{#2}%
    \phantomsection\label{#1}\endgroup
}
\makeatother

\newcounter{dummy} 
\numberwithin{dummy}{section}

\newtheorem*{theorem*}{Theorem}

\newtheorem{counterexample}[dummy]{Counterexample}
\newtheorem*{counterexample*}{Counterexample}
\newtheorem*{definition*}{Definition}

\newcounter{hackcount}

\title[Erratum]{Comments on ``$\Oh{m\cdot n}$~algorithms for the recognition and isomorphism problems on circular-arc graphs''}

\author[T.~Krawczyk]{Tomasz Krawczyk}
\address[T.~Krawczyk]{Faculty of Mathematics and Information Science, Warsaw University of Technology, Warsaw, Poland}
\email{tomasz.krawczyk@pw.edu.pl}
\date{}

\begin{document}

\thispagestyle{empty}

\begin{abstract}
In the work [\emph{$\Oh{m\cdot n}$~algorithms for the recognition and isomorphism problems on circular-arc graphs}, \emph{SIAM J. Comput. 24(3), 411--439, (1995)}], Wen-Lian Hsu claims 
three results concerning the class of circular-arc graphs:
\begin{itemize}
 \item the design of so-called \emph{decomposition trees} that represent the structure of all normalized intersection models of circular-arc graphs,
 \item an $\Oh{nm}$ recognition algorithm for circular-arc graphs,
 \item an $\Oh{nm}$ isomorphism algorithm for circular-arc graphs.
\end{itemize}
In [\emph{Discrete Math. Theor. Comput. Sci., 15(1), 157--182, 2013}] Curtis, Lin, McConnell, Nussbaum, Soulignac, Spinrad, and Szwarcfiter showed that Hsu's isomorphism algorithm is incorrect. 
In this note, we show that the other two results -- namely, the construction of decomposition trees and the recognition algorithm -- are also flawed.
\end{abstract}

\maketitle

\section{Introduction}

Given a family $\mathcal{F}$ of sets, the \emph{intersection graph} of $\mathcal{F}$ is obtained by representing every set in $\mathcal{F}$ by a vertex and connecting two vertices by an edge if and only if their corresponding sets have non-empty intersection. 
A~graph~$G$ is:
\begin{itemize}
 \item a \emph{circular-arc graph} if $G$ is the intersection graph of arcs of a circle,
 \item a \emph{circle graph} if $G$ is the intersection graph of chords of a circle,
 \item a \emph{permutation graph} if $G$ is the intersection graphs of chords 
 spanned between some two disjoint arcs of a circle.
\end{itemize}

In the work [\emph{\mbox{$\Oh{m\cdot n}$~algorithms} for the recognition and isomorphism problems on circular-arc graphs}, \emph{SIAM J. Comput. 24(3), 411--439, (1995)}] \cite{Hsu95}, Wen-Lian Hsu presents
three results concerning the class of circular-arc graphs.
The first of these results claims a construction of a \emph{decomposition tree} of a circular-arc graph, 
a data structure intended to represent the set of all its \emph{normalized intersection models}
(an analogue of the PQ-tree for an interval graph).
\emph{Normalized intersection models} (formally defined in Section~\ref{sec:preliminaries}) of a~circular-arc graph reflect the neighborhood relation between its vertices and can be seen as its canonical representations; 
in particular, any intersection model can be made normalized by possibly extending some of its arcs.
Based on these decomposition trees, Hsu claims $\Oh{nm}$-algorithms for both the recognition and isomorphism problem for circular-arc graphs.

In this short note we show that the description of the normalized intersection models given by Hsu in~\cite{Hsu95} is not correct. 
Specifically, we show the following:
\begin{counterexample}
\label{count:main_counter}
There are circular-arc graphs whose all normalized models do not follow the description given by Hsu. 
\end{counterexample}
As a result, the decomposition trees developed by Hsu, along with his polynomial-time algorithms for recognizing and testing isomorphism of circular-arc graphs, are incorrect.
We note that Curtis, Lin, McConnell, Nussbaum, Soulignac, Spinrad, and Szwarcfiter ~\cite{counterex13} identified a different flaw in Hsu's isomorphism algorithm already in 2013.
The approach used by Hsu (described below) determines somehow the main steps 
that need to be taken to describe the structure of the normalized models of a circular-arc graph. 
In this paper, we also identify several errors in Hsu's work, showing that these steps were either partially or entirely incorrectly accomplished.




To describe the structure of the normalized models of a circular-arc graph Hsu uses a~method that has its roots in the work of Gallai.
In his seminal paper~\cite{Gal67} from 1967 Gallai introduced a data structure, called \emph{modular decomposition tree}, which keeps a track of all \emph{modules} of a simple graph~$G$.
We recall that a set $M \subseteq V(G)$ is a \emph{module} of~$G$ if every vertex from outside $M$ is either adjacent to all vertices in~$M$ or non-adjacent to all
vertices in~$M$.
In the same work Gallai described the structure of all transitive orientations of
a comparability graphs.
In particular, Gallai proved that:
\begin{description}
 \item[\namedlabel{prop:G1}{(G1)}] If $G$ is \emph{prime} (contains only trivial modules, $V(G)$ and $\{v\}$ for all $v \in V(G)$), 
 then $G$ has a unique transitive orientation (up to reversal).
 \item[\namedlabel{prop:G2}{(G2)}] Otherwise, the structure of all transitive orientations of $G$ can be described (and represented) by means of the modular decomposition tree of $G$.
\end{description}
The above-mentioned results of Gallai have turned out to be widely used to describe
the structure of some geometric intersection graphs, especially those related to partial orders.
Among them are, for example, permutation graphs~\cite{DM41}, interval graphs\footnote{A reader familiar with PQ-trees and modular decomposition trees might easily observe that the PQ-tree representing the structure of intersection models of an interval graph can be easily obtained from its modular decomposition tree.} \cite{BoothLueker76}, or trapezoid graphs~\cite{MaS94a}.
In particular, Properties~\ref{prop:G1} and~\ref{prop:G2} allow to show that for 
every graph~$G$ from each of those graph classes, if a certain graph~$G'$ associated with~$G$ is prime\footnote{$G'=G$ if $G$ is an interval or a permutation graph, $G'$ is so-called \emph{split} of $G$ if $G$ is a trapezoid graph.}, then $G$ has a unique intersection model (up to some normalizations and reflections), and otherwise, if~$G'$ is not prime,
the structure of all intersection models of $G$ can be represented by the modular decomposition tree of $G'$.
As for circular-arc graphs, the work \cite{Spin88} of Spinrad from 1988 is the first that allows to describe, in the way given above, the structure of the normalized models of certain graphs in this class, namely 
those that are \emph{co-bipartite} (whose vertex set can be partitioned into two cliques). 
See~\cite{Kra24} where we briefly outline the influence of Spinrad's work~\cite{Spin88} 
on the development of the method applied in~\cite{Hsu95} and in~\cite{Kra24}.

In~\cite{Hsu95} Hsu tries to use Gallai's framework to describe the structure of the normalized models of any circular-arc graph~$G$.
For this purpose, Hsu transforms every normalized model~$R$ of~$G$ to a chord model~$D$ by 
converting each arc~$R(v)$ into a chord~$D(v)$ with the same endpoints as~$R(v)$, for every $v \in V(G)$.
The crucial observation of Hsu is that the resulting chord model $D$ represents the same circle graph~$G_c$,  
regardless of a normalized model $R$ of $G$.
Then, Hsu tries to describe the structure of the chord models of $G_c$ that correspond to the normalized models of $G$ -- such models of $G_{c}$ are called \emph{conformal} to~$G$ (shortly, conformal).
To this end, Hsu attempts to prove the properties similar to~\ref{prop:G1} and~\ref{prop:G2}:
\begin{description}
 \item[\namedlabel{prop:H1}{(H1)}] If $G_c$ is prime, 
 then $G_c$ has a unique conformal model (up to reflection).
 \item[\namedlabel{prop:H2}{(H2)}] Otherwise, the structure of the conformal models of $G_c$ can be described by means of the modular decomposition tree of $G_c$.
\end{description}
When $G$ is co-biparite, $G_c$ is a permutation graph, and hence Property~\ref{prop:H1} follows by~\ref{prop:G1} and Property~\ref{prop:H2} follows by~\ref{prop:G2} and by the work~\cite{Spin88} of Spinrad -- see~\cite{Kra24} for more details.

We want to emphasize that Property~\ref{prop:H1} is crucial for the method as
the structure of the conformal models is ``build'' upon the unique conformal models of prime subgraphs of~$G_c$ (the role of~\ref{prop:H1} is similar to the role of the fact that every prime subgraph has a unique transitive orientation in the description of the structure of all transitive orientations of a comparability graph).
In~Subsection~\ref{subsec:prime_overlap_graphs_and_conformal_models} we show that the ``proof'' of~\ref{prop:H1} proposed by Hsu (Theorem~5.7 in~\cite{Hsu95}) is not correct --
it is based on two claims, both of which are shown to be false in Section~\ref{sec:prime_case}.
Nevertheless, the statement of Property~\ref{prop:H1} is true, which follows from the proof presented in~\cite{Kra24}.

To accomplish~\ref{prop:H2} Hsu divides the description of the structure of the conformal models of $G_c$ into three cases: when $G_c$ is disconnected, when the complement $\overline{G_c}$ of $G_c$ is disconnected, and when both $G_c$ and $\overline{G_c}$ are connected.
Hsu's work~\cite{Hsu95} correctly deals with the case when $\overline{G_c}$ is disconnected (then, each component of $\overline{G_c}$ induces a co-bipartite circular-arc graph), it is incorrect when both $G_c$
and $\overline{G_c}$ are connected, and it is incomplete (and also incorrect) when $G_c$ is disconnected.
 
For the case when both $G_c$ and $\overline{G_c}$ are connected, 
Hsu first ``proves'' Property~\ref{prop:H1}.
Next, Hsu tries to partition the set $V(G_c)$ into so-called \emph{consistent modules} of $G_c$, which were supposed to satisfy the following property:
\begin{description}
 \item[\namedlabel{prop:H3}{(H3)}] For every conformal model $D$ and every consistent module $M$ of $G_c$ there are
 two disjoint arcs $A_M$ and $B_M$ of the circle such that each arc $A_M$ and $B_M$ contains one endpoint of every chord from $D(M)$ and no endpoint of any other chord from $D(V \setminus M)$ (it means, in particular, that $G_c[M]$ is a permutation graph).
Moreover, there is a unique circular order (up to reflection) in which the arcs $A_M$ and $B_M$ associated with distinct consistent modules $M$ of~$G_c$ may occur around the circle in all conformal models of~$G_c$.
\end{description}
In Subsection~\ref{subsec:consistent_modules} we show that the ``consistent modules'' determined by Hsu do not satisfy Property~\ref{prop:H3}.
This enables us, in particular, to construct a circular-arc graph claimed by Counterexample~\ref{count:main_counter}. 
Hence, the theorems from~\cite{Hsu95} ``proving'' Property~\ref{prop:H3} for the ``consistent modules'' determined by Hsu are also not correct -- see Subsection \ref{subsec:consistent_modules} for more details.

The description of Hsu is also not correct in the case when $G_c$ is disconnected
(the errors mentioned above are transferred to the description of the conformal models of some components of $G_c$). 
Moreover, this part of Hsu's work is missing many important ingredients (the consistent modules are not defined for all components of $G_c$) -- we refer to Section~\ref{sec:parallel_case} for more details.

Although Hsu's work \cite{Hsu95} is flawed, the approach taken in it is appropriate.
In~\cite{Kra24} we assume a different definition of conformal models, which allows us to describe the structure of all normalized models of a circular-arc graph using the same method.
We want to mention that, despite the errors, Hsu's work had a great influence on the work done in~\cite{Kra24} -- we refer to~\cite{Kra24} when we mention the ideas of Hsu used in this work.
In particular, Hsu's ideas have significantly contributed to understanding the structure of the normalized models, even in cases where bugs were identified.

\section{Preliminaries}
\label{sec:preliminaries}
In this section, we introduce the notation from Hsu's work~\cite{Hsu95} that is needed to present counterexamples to the aforementioned (false) claims made by Hsu.

For a simple graph $G$, by $V(G)$ and $E(G)$ we denote the vertex set and the edge set of $G$, respectively.
For $v$ in $V(G)$, we denote by $N(v)$ (or $N_G(v)$) the set consisting of $v$ and all vertices adjacent to $v$ in $G$.
Vertices $v_1$ and $v_2$ in $G$ are said to form a \emph{similar pair} if 
$N(v_1) \setminus \{v_1\} = N(v_2) \setminus \{v_2\}$.
A vertex $v$ in $G$ is said to be a \emph{D-vertex} (short for \emph{dominating})
if $N(v) = V(G)$.
For a set $M \subseteq V(G)$ by $G[M]$ we denote the subgraph of~$G$ induced by the set $M$.

\subsection{The modular decomposition and the join decomposition}
\label{sec:modular-and-join-decompositions}

\subsubsection{Modular decomposition}
Let $G$ be a graph.
A \emph{module} $M$ of $G$ is a subset of $V(G)$ such that between $M$ and each vertex $v$
in $V(G)\setminus M$ there are either no edges or all possible edges. 
Module $M$ is \emph{connected} iff $G[M]$
is connected. 
Module $M$ is \emph{complement connected} iff the complement of $G[M]$ is connected. 
An unconnected module is called a \emph{parallel module}. 
A~connected module whose complement is unconnected is called a \emph{series module}. 
A~module that is both connected and complement connected is called a \emph{neighborhood module}. 
A~module $M$ of size greater than $1$ is \emph{nontrivial}
if $M \neq V(G)$. 
A~graph $G$ is \emph{s-inseparable} if it does not contain any nontrivial module.
See page 416, lines 24 to 30.
A~graph $G$ which is s-inseparable is also called \emph{prime}.

By definition, every module is exactly one of these three types. 
The modular decomposition starts with the module $V = V(G)$
and creates a corresponding root node in the decomposition tree labelled with S, P, or N,
depending on the type of $V$. 
Then it decomposes this module into components 
$M_1, M_2,\ldots, M_k$ using one of the following rules:
\begin{itemize}
 \item if $V$ is a series module, then $M_1,\ldots,M_k$ are the connected components of $\overline{G}$,
 \item if $V$ is a parallel module, then $M_1,\ldots,M_k$ are the connected components of $G$,
 \item if $V$ is a neighborhood module, then $M_1,\ldots,M_k$ are the maximal modules of $G$ different than $V$.
\end{itemize}
For each $M_i$, find the modular decomposition tree of~$G[M_i]$ and make its root node a son of the root node representing~$V(G)$. 
The resulting tree is called a \emph{modular decomposition tree} of~$G$; it is proven in~\cite{Gal67} that the modular decomposition tree of~$G$ is unique.
See page~416, lines -18 to -7.

\subsubsection{Join decomposition}
A graph $G$ is said to have a \emph{join} if $V$ can be partitioned into $V_0$, $V_1$,
$V_2$, and $V_3$ with $V_0 \cup V_1 \geq 2$ and $V_2 \cup V_3 \geq 2$ 
such that every possible edge exists between $V_1,V_2$ 
and no edge exists between $V_0, V_2 \cup V_3$, 
or between $V_0 \cup V_1,V_3$. 
In this case, we say $G$ is \emph{decomposable into $H_1$ and $H_2$}
where $H_1$ is the induced subgraph of $G[V_0 \cup V_1]$ with an
extra vertex $v_1$ adjacent to every vertex in $V_1$ 
and $H_2$ is the induced subgraph of $G[V_2 \cup V_3]$ with
an extra vertex $v_2$ adjacent to every vertex in $V_2$. 
If the above holds, then $(V_0,V_1,V_2,V_3)$ is a \emph{join} of $G_c$ and
$H_1$ and $H_2$ is a \emph{decomposition} of $G_c$ \emph{induced by the join} $(V_0,V_1,V_2,V_3)$.
A graph $G$ is said to be \emph{j-inseparable} if it
does not contain any join.
See page 417, lines -7 to -1.

\subsection{Circular-arc graphs and its normalized models}
\begin{figure}[htp!]
\begin{tikzpicture}[scale=0.5]
\coordinate (center) at (0,0) {};
\coordinate (v) at ($(center)+(90:2cm)$) {};
\coordinate (u) at ($(center)+(270:2cm)$) {};

\coordinate (lv) at ($(center)+(90:2.6cm)$) {};
\coordinate (lu) at ($(center)+(270:2.6cm)$) {};

\tikzstyle{every node}=[inner sep=1pt]
\begin{scriptsize}
\node at (lv) {$u_1$};
\node at (lu) {$u_2$};
\end{scriptsize}

\draw[thick] ([shift=(30:2.1cm)]0,0) arc (30:150:2.1cm);
\draw[thick] ([shift=(210:2.1cm)]0,0) arc (210:330:2.1cm);

\draw[thick, white] (-3.0,-3)--(-3.0,-2.8);
\draw[thick, white] (3.0,3)--(3.0,2.8);

\end{tikzpicture}
\begin{tikzpicture}[scale=0.5]
\coordinate (center) at (0,0) {};
\coordinate (v) at ($(center)+(90:2cm)$) {};
\coordinate (u) at ($(center)+(270:2cm)$) {};

\coordinate (lv) at ($(center)+(90:2.6cm)$) {};
\coordinate (lu) at ($(center)+(90:1.25cm)$) {};

\tikzstyle{every node}=[inner sep=1pt]
\begin{scriptsize}
\node at (lv) {$u_1$};
\node at (lu) {$u_2$};
\end{scriptsize}

\draw[thick] ([shift=(0:2.1cm)]0,0) arc (0:180:2.1cm);
\draw[thick] ([shift=(45:1.8cm)]0,0) arc (45:135:1.8cm);

\draw[thick, white] (-3.0,-3)--(-3.0,-2.8);
\draw[thick, white] (3.0,3)--(3.0,2.8);

\end{tikzpicture}
\begin{tikzpicture}[scale=0.5]
\coordinate (center) at (0,0) {};
\coordinate (v) at ($(center)+(90:2cm)$) {};
\coordinate (u) at ($(center)+(270:2cm)$) {};

\coordinate (lu) at ($(center)+(90:2.6cm)$) {};
\coordinate (lv) at ($(center)+(90:1.25cm)$) {};

\tikzstyle{every node}=[inner sep=1pt]
\begin{scriptsize}
\node at (lv) {$u_1$};
\node at (lu) {$u_2$};
\end{scriptsize}

\draw[thick] ([shift=(0:2.1cm)]0,0) arc (0:180:2.1cm);
\draw[thick] ([shift=(45:1.8cm)]0,0) arc (45:135:1.8cm);

\draw[thick, white] (-3.0,-3)--(-3.0,-2.8);
\draw[thick, white] (3.0,3)--(3.0,2.8);

\end{tikzpicture}
\begin{tikzpicture}[scale=0.5]
\coordinate (center) at (0,0) {};
\coordinate (v) at ($(center)+(90:2cm)$) {};
\coordinate (u) at ($(center)+(270:2cm)$) {};

\coordinate (lv) at ($(center)+(90:2.6cm)$) {};
\coordinate (lu) at ($(center)+(270:2.4cm)$) {};

\tikzstyle{every node}=[inner sep=1pt]
\begin{scriptsize}
\node at (lv) {$u_1$};
\node at (lu) {$u_2$};
\end{scriptsize}

\draw[thick] ([shift=(-20:2.1cm)]0,0) arc (-20:200:2.1cm);
\draw[thick] ([shift=(160:1.9cm)]0,0) arc (160:380:1.9cm);

\draw[thick, white] (-3.0,-3)--(-3.0,-2.8);
\draw[thick, white] (3.0,3)--(3.0,2.8);

\end{tikzpicture}
\begin{tikzpicture}[scale=0.5]
\coordinate (center) at (0,0) {};
\coordinate (v) at ($(center)+(90:2cm)$) {};
\coordinate (u) at ($(center)+(270:2cm)$) {};

\coordinate (lv) at ($(center)+(90:2.6cm)$) {};
\coordinate (lu) at ($(center)+(270:2.4cm)$) {};

\tikzstyle{every node}=[inner sep=1pt]
\begin{scriptsize}
\node at (lv) {$u_1$};
\node at (lu) {$u_2$};
\end{scriptsize}
\draw[thick] ([shift=(70:2.1cm)]0,0) arc (70:200:2.1cm);
\draw[thick] ([shift=(160:1.9cm)]0,0) arc (160:290:1.9cm);

\draw[thick, white] (-3.0,-3)--(-3.0,-2.8);
\draw[thick, white] (3.0,3)--(3.0,2.8);
\end{tikzpicture}

\caption{\label{fig:arcs_relation} From left to right:
$u_1$ and $u_2$ are independent, $u_1$ contains~$u_2$, $u_1$ is contained in~$u_2$,
$u_1$ and $u_2$ cover the circle, and $u_1$ and $u_2$ strictly overlap.
}

\end{figure}
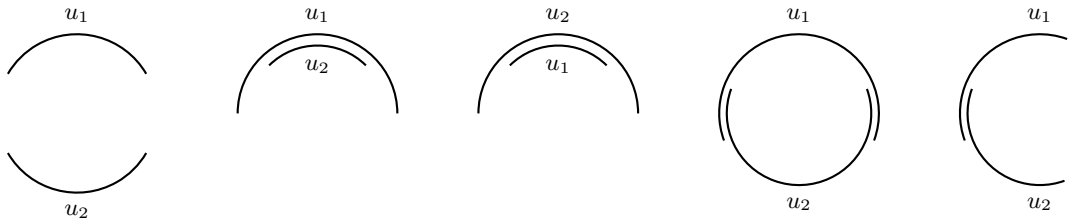

Given a circular-arc model~$R$ of a circular-arc graph~$G$ such that all arcs in~$R$ have pairwise distinct endpoints, Hsu distinguishes five possible relationships on a pair of arcs $u_1$ and $u_2$ in $R$, 
shown in Figure~\ref{fig:arcs_relation}.
See page 413, lines~-11 to~-3, in~\cite{Hsu95}.

Following Hsu (see page 414, lines 3 to 8), two vertices $v_1$ and $v_2$ in a graph $G$ are said to be:
\begin{enumerate}
\item \emph{independent} if $v_1$ is not adjacent to $v_2$.
\item \emph{strictly adjacent} if $v_1$ is adjacent to $v_2$ but neither $N(v_1)$ nor $N(v_2)$ is contained in the other.
\item \emph{strongly adjacent} if $v_1$ and $v_2$ are strictly adjacent 
but every $w$ in $V(G) \setminus N(v_1)$ satisfies $N(w) \subseteq N(v_2)$ and every $w'$ in $V(G)\setminus N(v_2)$ satisfies   $N(w') \subseteq N(v_1)$.
\item \emph{similar} if $N(v_1)\setminus \{v_1\} = N(v_2) \setminus \{v_2\}$.
\end{enumerate}

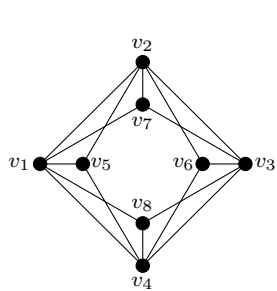
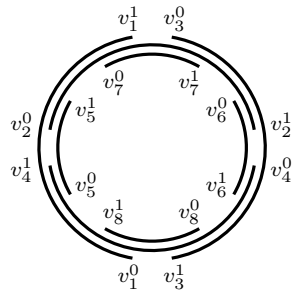
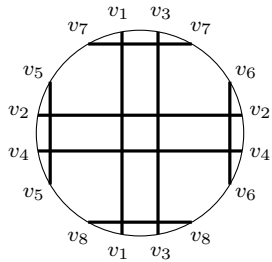
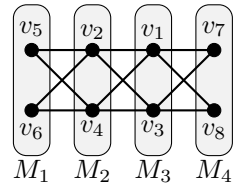
\begin{figure}[htp!]
\centering \small
\begin{subfigure}[b]{0.45\linewidth}
    \centering
\begin{tikzpicture}[scale=0.45,>=latex]
\coordinate (center) at (0.0,0.0) {};


\coordinate (u1) at ($(center)+(180:1.75cm)$) {};
\coordinate (v1) at ($(center)+(180:3cm)$) {};
\coordinate (lu1) at ($(center)+(180:1.2cm)$) {};
\coordinate (lv1) at ($(center)+(180:3.6cm)$) {};

\coordinate (u2) at ($(center)+(90:1.75cm)$) {};
\coordinate (v2) at ($(center)+(90:3cm)$) {};
\coordinate (lu2) at ($(center)+(90:1.2cm)$) {};
\coordinate (lv2) at ($(center)+(90:3.5cm)$) {};

\coordinate (u3) at ($(center)+(0:1.75cm)$) {};
\coordinate (v3) at ($(center)+(0:3cm)$) {};
\coordinate (lu3) at ($(center)+(0:1.2cm)$) {};
\coordinate (lv3) at ($(center)+(0:3.6cm)$) {};

\coordinate (u4) at ($(center)+(270:1.75cm)$) {};
\coordinate (v4) at ($(center)+(270:3cm)$) {};
\coordinate (lu4) at ($(center)+(270:1.2cm)$) {};
\coordinate (lv4) at ($(center)+(270:3.5cm)$) {};

\tikzstyle{every node}=[circle,minimum size=5pt,inner sep=0pt,draw,fill]
\node at (u1) {};
\node at (v1) {};
\node at (u2) {};
\node at (v2) {};
\node at (u3) {};
\node at (v3) {};
\node at (u4) {};
\node at (v4) {};

\tikzstyle{every node}=[inner sep=1pt]
\begin{tiny}
\node at (lv1) {$v_1$};
\node at (lu1) {$v_5$};

\node at (lv2) {$v_2$};
\node at (lu2) {$v_7$};

\node at (lv3) {$v_3$};
\node at (lu3) {$v_6$};

\node at (lv4) {$v_4$};
\node at (lu4) {$v_8$};
\end{tiny}

\draw[-] (u1)--(v1);
\draw[-] (u1)--(v2);
\draw[-] (u1)--(v4);

\draw[-] (u2)--(v2);
\draw[-] (u2)--(v1);
\draw[-] (u2)--(v3);

\draw[-] (u3)--(v3);
\draw[-] (u3)--(v2);
\draw[-] (u3)--(v4);

\draw[-] (u4)--(v4);
\draw[-] (u4)--(v1);
\draw[-] (u4)--(v3);

\draw[-] (v1)--(v2);
\draw[-] (v2)--(v3);
\draw[-] (v3)--(v4);
\draw[-] (v4)--(v1);

\draw[white] (-3.7,-3.7)--(-3.7,-3.3);
\draw[white] (3.7,4)--(3.7,3.7);
\end{tikzpicture}
\caption{A circular-arc graph $G$}
  \label{fig:counter_circular_arc_graph}
\end{subfigure}
  \,
\begin{subfigure}[b]{0.45\linewidth}
    \centering
\begin{tikzpicture}[scale=0.62,>=latex]
\coordinate (center) at (0.0,0.0) {};

\coordinate (lu03) at ($(center)+(30:1.6cm)$) {};
\coordinate (lu13) at ($(center)+(-30:1.6cm)$) {};
\draw[very thick,-] ([shift=(-30:2cm)]0,0) arc (-30:30:2cm);

\coordinate (lu12) at ($(center)+(60:1.6cm)$) {};
\coordinate (lu02) at ($(center)+(120:1.6cm)$) {};
\draw[very thick,-] ([shift=(60:2cm)]0,0) arc (60:120:2cm);

\coordinate (lu11) at ($(center)+(150:1.6cm)$) {};
\coordinate (lu01) at ($(center)+(210:1.6cm)$) {};
\draw[very thick,-] ([shift=(150:2cm)]0,0) arc (150:210:2cm);

\coordinate (lu14) at ($(center)+(240:1.6cm)$) {};
\coordinate (lu04) at ($(center)+(300:1.6cm)$) {};
\draw[very thick,-] ([shift=(240:2cm)]0,0) arc (240:300:2cm);

\coordinate (lv12) at ($(center)+(10:2.8cm)$) {};
\coordinate (lv02) at ($(center)+(170:2.8cm)$) {};
\draw[very thick,-] ([shift=(10:2.2cm)]0,0) arc (10:170:2.2cm);

\coordinate (lv14) at ($(center)+(190:2.8cm)$) {};
\coordinate (lv04) at ($(center)+(350:2.8cm)$) {};
\draw[very thick,-] ([shift=(190:2.2cm)]0,0) arc (190:350:2.2cm);

\coordinate (lv11) at ($(center)+(100:2.8cm)$) {};
\coordinate (lv01) at ($(center)+(260:2.8cm)$) {};
\draw[very thick,-] ([shift=(100:2.4cm)]0,0) arc (100:260:2.4cm);

\coordinate (lv13) at ($(center)+(-80:2.8cm)$) {};
\coordinate (lv03) at ($(center)+(80:2.8cm)$) {};
\draw[very thick,-] ([shift=(-80:2.4cm)]0,0) arc (-80:80:2.4cm);

\tikzstyle{every node}=[inner sep=1pt]
\begin{tiny}
\node at (lv01) {$v^0_1$};
\node at (lv11) {$v^1_1$};

\node at (lv02) {$v^0_2$};
\node at (lv12) {$v^1_2$};

\node at (lv03) {$v^0_3$};
\node at (lv13) {$v^1_3$};

\node at (lv04) {$v^0_4$};
\node at (lv14) {$v^1_4$};

\node at (lu01) {$v^0_5$};
\node at (lu11) {$v^1_5$};

\node at (lu02) {$v^0_7$};
\node at (lu12) {$v^1_7$};

\node at (lu03) {$v^0_6$};
\node at (lu13) {$v^1_6$};

\node at (lu04) {$v^0_8$};
\node at (lu14) {$v^1_8$};
\end{tiny}

\draw[white] (-3,-3)--(-3,-2.7);
\draw[white] (3,3)--(3,2.7);
\end{tikzpicture}
\caption{Normalized model $R$ of $G$}
  \label{fig:counter_normalized_model}
\end{subfigure}

\begin{subfigure}[b]{0.45\linewidth}
    \centering
\begin{tikzpicture}[scale=0.62,>=latex]
\coordinate (center) at (0.0,0.0) {};
\draw (center) circle (2.2cm);

\coordinate (lu03) at ($(center)+(30:2.6cm)$) {};
\coordinate (lu13) at ($(center)+(-30:2.6cm)$) {};
\draw[very thick,-] ($(center)+(-30:2.2cm)$)--($(center)+(30:2.2cm)$);

\coordinate (lu12) at ($(center)+(60:2.6cm)$) {};
\coordinate (lu02) at ($(center)+(120:2.6cm)$) {};
\draw[very thick,-] ($(center)+(60:2.2cm)$)--($(center)+(120:2.2cm)$);

\coordinate (lu11) at ($(center)+(150:2.6cm)$) {};
\coordinate (lu01) at ($(center)+(210:2.6cm)$) {};
\draw[very thick,-] ($(center)+(150:2.2cm)$)--($(center)+(210:2.2cm)$);

\coordinate (lu14) at ($(center)+(240:2.6cm)$) {};
\coordinate (lu04) at ($(center)+(300:2.6cm)$) {};
\draw[very thick,-] ($(center)+(240:2.2cm)$)--($(center)+(300:2.2cm)$);

\coordinate (lv12) at ($(center)+(10:2.6cm)$) {};
\coordinate (lv02) at ($(center)+(170:2.6cm)$) {};
\draw[very thick,-] ($(center)+(10:2.2cm)$)--($(center)+(170:2.2cm)$);

\coordinate (lv14) at ($(center)+(190:2.6cm)$) {};
\coordinate (lv04) at ($(center)+(350:2.6cm)$) {};
\draw[very thick,-] ($(center)+(190:2.2cm)$)--($(center)+(350:2.2cm)$);

\coordinate (lv11) at ($(center)+(100:2.6cm)$) {};
\coordinate (lv01) at ($(center)+(260:2.6cm)$) {};
\draw[very thick,-] ($(center)+(100:2.2cm)$)--($(center)+(260:2.2cm)$);

\coordinate (lv13) at ($(center)+(-80:2.6cm)$) {};
\coordinate (lv03) at ($(center)+(80:2.6cm)$) {};
\draw[very thick,-] ($(center)+(-80:2.2cm)$)--($(center)+(80:2.2cm)$);

\tikzstyle{every node}=[inner sep=1pt]
\begin{tiny}
\node at (lv01) {$v_1$};
\node at (lv11) {$v_1$};

\node at (lv02) {$v_2$};
\node at (lv12) {$v_2$};

\node at (lv03) {$v_3$};
\node at (lv13) {$v_3$};

\node at (lv04) {$v_4$};
\node at (lv14) {$v_4$};

\node at (lu01) {$v_5$};
\node at (lu11) {$v_5$};

\node at (lu02) {$v_7$};
\node at (lu12) {$v_7$};

\node at (lu03) {$v_6$};
\node at (lu13) {$v_6$};

\node at (lu04) {$v_8$};
\node at (lu14) {$v_8$};
\end{tiny}

\draw[white] (-3,-3)--(-3,-2.7);
\draw[white] (3,3)--(3,2.7);

\end{tikzpicture}
\caption{Associated chord model $D$ of $G_c$}
  \label{fig:counter_circle_graph}
\end{subfigure}
\
\begin{subfigure}[b]{0.45\linewidth}
    \centering
\begin{tikzpicture}[xscale=0.4,yscale=0.4]
\coordinate (lu1) at (-2,1.7) {};
\coordinate (u1) at (-2,1) {};
\coordinate (u3) at (-2,-1) {};
\coordinate (lu3) at (-2,-1.7) {};

\coordinate (lv2) at (0,1.6) {};
\coordinate (v2) at (0,1) {};
\coordinate (v4) at (0,-1) {};
\coordinate (lv4) at (0,-1.6) {};

\coordinate (lv1) at (2,1.6) {};
\coordinate (v1) at (2,1) {};
\coordinate (v3) at (2,-1) {};
\coordinate (lv3) at (2,-1.6) {};

\coordinate (lu2) at (4,1.7) {};
\coordinate (u2) at (4,1) {};
\coordinate (u4) at (4,-1) {};
\coordinate (lu4) at (4,-1.7) {};

\coordinate (lM1) at (-2,-3) {};  
\coordinate (lM2) at (0,-3) {};  
\coordinate (lM3) at (2,-3) {};  
\coordinate (lM4) at (4,-3) {};  
\draw[rounded corners=5, fill=gray!10] (-2.6,-2.5)--(-2.6,2.5) -- (-1.4,2.5) -- (-1.4,-2.5)--cycle;
\draw[rounded corners=5, fill=gray!10] (-0.6,-2.5)--(-0.6,2.5) -- (0.6,2.5) -- (0.6,-2.5)--cycle;
\draw[rounded corners=5, fill=gray!10] (1.4,-2.5)--(1.4,2.5) -- (2.6,2.5) -- (2.6,-2.5)--cycle;
\draw[rounded corners=5, fill=gray!10] (3.4,-2.5)--(3.4,2.5) -- (4.6,2.5) -- (4.6,-2.5)--cycle;

\tikzstyle{every node}=[circle,minimum size=5pt,inner sep=0pt,draw,fill]
\node at (u1) {};
\node at (u3) {};
\node at (u2) {};
\node at (u4) {};
\node at (v1) {};
\node at (v2) {};
\node at (v3) {};
\node at (v4) {};

\tikzstyle{every node}=[inner sep=1pt]

\path (u1) edge[thick] (v2);
\path (u1) edge[thick] (v4);
\path (u3) edge[thick] (v2);
\path (u3) edge[thick] (v4);

\path (v2) edge[thick] (v1);
\path (v2) edge[thick] (v3);
\path (v4) edge[thick] (v1);
\path (v4) edge[thick] (v3);

\path (v1) edge[thick] (u2);
\path (v1) edge[thick] (u4);
\path (v3) edge[thick] (u2);
\path (v3) edge[thick] (u4);

\begin{footnotesize}
\tikzstyle{every node}=[inner sep=2pt]
\node at (lM1) {$M_1$};
\node at (lM2) {$M_2$};
\node at (lM3) {$M_3$};
\node at (lM4) {$M_4$};

\node at (lu1) {$v_5$};
\node at (lu2) {$v_7$};
\node at (lu3) {$v_6$};
\node at (lu4) {$v_8$};

\node at (lv1) {$v_1$};
\node at (lv2) {$v_2$};
\node at (lv3) {$v_3$};
\node at (lv4) {$v_4$};

\end{footnotesize}

\draw[white] (-3.2,-4.6)--(-3.2,-2.7);
\draw[white] (5,3)--(5,2.7);
\end{tikzpicture}
\caption{Graph $G_c$ and its maximal non-trivial modules $M_1,M_2,M_3,M_4$}
  \label{fig:counter_overlap_graph}
\end{subfigure}
 \caption{Counterexample~\ref{count:main_counter}: $V(G_c)$ is a neighbourhood module, $M_1,M_2,M_3,M_4$ are all parallel children of $V(G_c)$
in the modular decomposition tree of $G_c$, and $M_1$ and $M_4$ are not consistent. 
}
\label{fig:ca_graph_transformation}
\end{figure}

\noindent \textbf{Definition [page 414, lines -14 to -5]}
\emph{Let $G$ be a circular-arc graph containing neither similar pairs nor D-vertices.
A circular-arc model $R$ for $G$ is said to be a \emph{normalized model (N-model)} 
if every pair of arcs $u_1$ and $u_2$ in $R$ and their corresponding vertices $v_1$ and $v_2$ in $G$ satisfy the following conditions:
\begin{enumerate}
\item $u_1$ is independent of $u_2$ $\iff$ $v_1$ is not adjacent to $v_2$.
\item $u_1$ is contained in $u_2$ $\iff$ $N(v_1) \subseteq N(v_2)$.
\item $u_1$ strictly overlaps with $u_2$ $\iff$ $v_1$ and $v_2$ are strictly but not strongly adjacent.
\item $u_1$ and $u_2$ cover the circle $\iff$ $v_1$ and $v_2$ are strongly adjacent.
\end{enumerate}}
Hsu showed that, if $G$ contains neither similar pairs nor D-vertices, every circular-arc model of $G$ can be turned into a normalized one by possibly extending some of its arcs.
Figure~\ref{fig:ca_graph_transformation}.(A) shows an exemplary circular-arc graph and Figure~\ref{fig:ca_graph_transformation}.(B) presents one of its normalized model.

\noindent \textbf{Definition [page 420, lines 11 to 12]}
\emph{Associate with each graph $G$ the graph $G_c$ that has the same vertex set as $G$
such that two vertices in $G_c$ are adjacent iff they are strictly but not strongly adjacent in $G$.}

\emph{Consider a circular-arc graph $G$ with an N-model $R$. 
Associate with each arc in $R$ a~chord that connects its two endpoints. 
The resulting chord model $D$ is called an \emph{associated chord
model} for $G$. 
Since two arcs cross in $R$ exactly when their corresponding chords in $D$ cross,
this chord model gives rise to the circle graph $G_c$. [...]
Although there can be many associated chord models for $G$, they all give rise to the unique circle graph $G_c$ defined above.}
See page 420, lines -15 to -10, and Figure~\ref{fig:ca_graph_transformation} for an illustration.

Next, Hsu lists some properties of the chord models $D$ of~$G_c$ obtained from the normalized models $R$ of~$G$.
For every vertex $u \in V(G)$ let $I_{u}$ be the set of vertices not adjacent to $u$ in $G_c$, and let
$$
\begin{array}{rcl}
L_u &=& \{ v \in I_u: \quad
\text{$N(v) \subsetneq N(u)$} \quad \text{or} \quad \text{$u$ and $v$ are strongly adjacent}\}, 
\\
R_u &=& \{ v \in I_u:
\quad \text{$u$ and $v$ are disjoint} \quad \text{or} \quad \text{$N(u) \subsetneq N(v)$}\}.
\end{array}
$$
Hsu notes that for every $u \in V$ the chords representing the vertices from the set $L_u$ are on one side of $D(u)$ and the chords representing the vertices from the set $R_u$ are on the other side of $D(u)$.
Based on this observation, Hsu assumes the following definition:

\smallskip
\noindent \textbf{Definition [page 424, lines 18 to 20]}
\emph{A chord model $D$ of~$G_c$ is \emph{conformal} if for every vertex $u$ of $G$ the chords associated with vertices in $L_u$ are on one side of the chord of $D(u)$ and those associated with vertices in $R_u$ are on the other side of the chord of $D(u)$.}

\smallskip
Then, Hsu proves the following:

\smallskip
\noindent \textbf{Theorem 5.6 [page 424, lines -18 to -17]}
\emph{Let $G$ be a circular-arc graph [...]. 
Then a model for $G_c$ is conformal if and only if it is a chord model corresponding to some normalized model for~$G$.}

\smallskip

In Hsu's work the description of the structure of the conformal models of $G_{c}$ is split into three parts, corresponding to the cases when $V(G_{c})$ is a series, neighbourhood, or parallel module in the modular decomposition tree of $G_{c}$.
Hsu's work~\cite{Hsu95} correctly deals with the case where $V(G_{c})$ is series, 
it is not correct when $V(G_c)$ is a neighborhood module, and is incomplete (and also not correct) when $V(G_c)$ is a parallel module.

\section{$V(G_c)$ is a neighbourhood module}
\label{sec:prime_case}

\subsection{Conformal models of prime graphs $G_c$}
\label{subsec:prime_overlap_graphs_and_conformal_models}
When $V(G_c)$ is a neighborhood module of~$G_c$, 
Hsu considers first the case when $G_c$ is s-inseparable 
(if $G_c$ is s-inseparable, then both $G_c$ and $\overline{G_c}$ are connected, and $V(G_c)$ is a neighbourhood module in $G_c$).
For this case Hsu ``proves'' the following:

\smallskip

\noindent \textbf{Theorem 5.7 [page 425, lines 16 to 17]}
\emph{Let $G$ be a circular-arc graph.
If $G_c$ is \mbox{s-inseparable}, then $G$ has a unique N-model.}
\smallskip

The ``proof'' of Theorem~5.7 is split into two cases depending on whether
$G_c$ is j-inseparable.
If $G_c$ is j-inseparable, then $G_c$ has a unique 
chord model~\cite{GSH89}, and hence $G$ has a unique \mbox{N-model}.

So, the interesting case is when $G_c$ is not j-inseparable, that is,
when $G_c$ has a join $(V_0, V_1, V_2, V_3)$.
First, Hsu observes that $V_i \neq \emptyset$ for $i=1,\ldots,4$ as $G_s$
is s-inseparable.
Then, the proof is based on two claims; we show that both of them are false.

First, the ``proof'' of Theorem 5.7 ``shows'' the following property of the graph~$G_c$ 
and the join $(V_0,V_1,V_2,V_3)$ of~$G_c$:

\smallskip

\noindent \textbf{Claim A [page~425, line -23 to -18]:} \emph{If $|V_1| > 0$ and there exists a vertex $s$ in $V_1$ that is adjacent to all other vertices in $V_1 \cup V_2$, then it is easy to verify that $G_c \setminus \{s\}$ is \mbox{$s$-inseparable}.
We can proceed with the construction below to show that $G_c \setminus \{s\}$ has a unique conformal model, and, furthermore, there is a unique way to insert the chord of $s$ into this model.
The argument is symmetric if such a vertex $s$ belongs to $V_2$.}

\medskip

\begin{figure}[htp!]
  \centering \small
  \begin{subfigure}[b]{0.35\linewidth}
    \centering
\begin{tikzpicture}[scale=0.7,>=latex]

\coordinate (center) at (0.0,0.0) {};

\draw[black,very thick,-] ([shift=(90:2.2cm)]0,0) arc (90:270:2.2cm);
\coordinate (lv1) at ($(center)+(90:2.5cm)$) {};
\coordinate (lv0) at ($(center)+(270:1.9cm)$) {};

\draw[green,very thick,-] ([shift=(60:2cm)]0,0) arc (60:120:2cm);
\coordinate (lv01) at ($(center)+(120:1.7cm)$) {};
\coordinate (lv11) at ($(center)+(60:1.7cm)$) {};
\draw[red,very thick,-] ([shift=(75:2.4cm)]0,0) arc (75:30:2.4cm);
\coordinate (lv02) at ($(center)+(75:2.7cm)$) {};
\coordinate (lv12) at ($(center)+(30:2.7cm)$) {};
\draw[blue,very thick,-] ([shift=(45:2cm)]0,0) arc (45:0:2cm);
\coordinate (lv03) at ($(center)+(45:1.7cm)$) {};
\coordinate (lv13) at ($(center)+(0:1.7cm)$) {};
\draw[green,very thick,-] ([shift=(165:2.4cm)]0,0) arc (165:375:2.4cm);
\coordinate (lv14) at ($(center)+(165:2.7cm)$) {};
\coordinate (lv04) at ($(center)+(375:2.7cm)$) {};

\draw[blue,very thick,-] ([shift=(180:2cm)]0,0) arc (180:135:2cm);
\coordinate (lv05) at ($(center)+(180:1.7cm)$) {};
\coordinate (lv15) at ($(center)+(135:1.7cm)$) {};
\draw[red,very thick,-] ([shift=(105:2.4cm)]0,0) arc (105:150:2.4cm);
\coordinate (lv06) at ($(center)+(150:2.7cm)$) {};
\coordinate (lv16) at ($(center)+(105:2.7cm)$) {};

\tikzstyle{every node}=[inner sep=1pt]
\begin{tiny}
\node at (lv0) {$s^0$};
\node at (lv1) {$s^1$};

\node at (lv01) {$s^0_1$};
\node at (lv11) {$s^1_1$};

\node at (lv02) {$s^0_2$};
\node at (lv12) {$s^1_2$};

\node at (lv03) {$s^0_3$};
\node at (lv13) {$s^1_3$};

\node at (lv04) {$s^0_4$};
\node at (lv14) {$s^1_4$};

\node at (lv05) {$s^0_5$};
\node at (lv15) {$s^1_5$};

\node at (lv06) {$s^0_6$};
\node at (lv16) {$s^1_6$};

\end{tiny}

\draw[white] (-2.8,-2.8)--(-2.8,-2.5);
\draw[white] (2.8,2.8)--(2.8,2.5);

\end{tikzpicture}
    \caption{N-model $R^s$ of~$G^s$}\label{fig:claim_A_counter_normalized_model_gadget}
  \end{subfigure}
  \,
  \begin{subfigure}[b]{0.3\linewidth}
    \centering
\begin{tikzpicture}[scale=0.7,>=latex]
\coordinate (center) at (0.0,0.0) {};
\draw (center) circle (2.2cm);

\draw[black,very thick,-] ($(center)+(90:2.2cm)$)--($(center)+(270:2.2cm)$);
\coordinate (lv1) at ($(center)+(90:2.45cm)$) {};
\coordinate (lv0) at ($(center)+(270:2.45cm)$) {};

\draw[green,very thick,-] ($(center)+(60:2.2cm)$)--($(center)+(120:2.2cm)$);
\coordinate (lv11) at ($(center)+(120:2.45cm)$) {};
\coordinate (lv01) at ($(center)+(60:2.45cm)$) {};

\draw[red,very thick,-] ($(center)+(75:2.2cm)$)--($(center)+(30:2.2cm)$);
\coordinate (lv02) at ($(center)+(75:2.45cm)$) {};
\coordinate (lv12) at ($(center)+(30:2.45cm)$) {};

\draw[blue,very thick,-] ($(center)+(45:2.2cm)$)--($(center)+(0:2.2cm)$);
\coordinate (lv03) at ($(center)+(45:2.45cm)$) {};
\coordinate (lv13) at ($(center)+(0:2.45cm)$) {};

\draw[green,very thick,-] ($(center)+(165:2.2cm)$)--($(center)+(375:2.2cm)$);
\coordinate (lv14) at ($(center)+(165:2.45cm)$) {};
\coordinate (lv04) at ($(center)+(375:2.45cm)$) {};

\draw[blue,very thick,-] ($(center)+(180:2.2cm)$)--($(center)+(135:2.2cm)$);
\coordinate (lv15) at ($(center)+(180:2.45cm)$) {};
\coordinate (lv05) at ($(center)+(135:2.45cm)$) {};

\draw[red,very thick,-] ($(center)+(105:2.2cm)$)--($(center)+(150:2.2cm)$);
\coordinate (lv06) at ($(center)+(150:2.45cm)$) {};
\coordinate (lv16) at ($(center)+(105:2.45cm)$) {};

\tikzstyle{every node}=[inner sep=1pt]
\begin{tiny}
\node at (lv0) {$s$};
\node at (lv1) {$s$};

\node at (lv01) {$s_1$};
\node at (lv11) {$s_1$};

\node at (lv02) {$s_2$};
\node at (lv12) {$s_2$};

\node at (lv03) {$s_3$};
\node at (lv13) {$s_3$};

\node at (lv04) {$s_4$};
\node at (lv14) {$s_4$};

\node at (lv05) {$s_5$};
\node at (lv15) {$s_5$};

\node at (lv06) {$s_6$};
\node at (lv16) {$s_6$};

\end{tiny}

\draw[white] (-2.8,-2.8)--(-2.8,-2.5);
\draw[white] (2.8,2.8)--(2.8,2.5);

\end{tikzpicture}
    \caption{Ass. chord model of~$G^s_c$}\label{fig:claim_A_chord_model_gadget}
  \end{subfigure}
\,
  \begin{subfigure}[b]{0.3\linewidth}
    \centering
\begin{tikzpicture}[scale=0.7,>=latex]

\draw (center) circle (2cm);

\draw[black,very thick,<-] ($(center)+(90:2cm)$)--($(center)+(270:2cm)$);
\coordinate (lv1) at ($(center)+(90:2.3cm)$) {};
\coordinate (lv0) at ($(center)+(270:2.3cm)$) {};

\draw[green,ultra thick,-] ([shift=(60:2cm)]0,0) arc (60:120:2cm);
\draw[red,ultra thick,-] ([shift=(60:2cm)]0,0) arc (60:30:2cm);
\draw[blue,ultra thick,-] ([shift=(30:2cm)]0,0) arc (30:0:2cm);
\draw[blue,ultra thick,-] ([shift=(150:2cm)]0,0) arc (150:180:2cm);
\draw[red,ultra thick,-] ([shift=(120:2cm)]0,0) arc (120:150:2cm);

\tikzstyle{every node}=[inner sep=1pt]
\begin{tiny}
\node at (lv0) {$s^0$};
\node at (lv1) {$s^1$};
\end{tiny}

\draw[white] (-2.8,-2.8)--(-2.8,-2.5);
\draw[white] (2.8,2.8)--(2.8,2.5);

\end{tikzpicture}
    \caption{Schematic view of~$R^v$}\label{fig:B_counter_normalized_model}
\end{subfigure}

  \vspace{0.2cm}

\begin{subfigure}[b]{0.5\linewidth}
\centering
\begin{tikzpicture}[scale=0.7,>=latex]

\draw (center) circle (2cm);

\draw[black,very thick,<-] ($(center)+(90:2cm)$)--($(center)+(270:2cm)$);
\coordinate (lv1) at ($(center)+(90:2.3cm)$) {};
\coordinate (lv0) at ($(center)+(270:2.3cm)$) {};
\draw[green,ultra  thick,-] ([shift=(80:2cm)]0,0) arc (80:100:2cm);
\draw[red,ultra thick,-] ([shift=(70:2cm)]0,0) arc (70:80:2cm);
\draw[blue,ultra thick,-] ([shift=(60:2cm)]0,0) arc (60:70:2cm);
\draw[red,ultra thick,-] ([shift=(100:2cm)]0,0) arc (100:110:2cm);
\draw[blue,ultra thick,-] ([shift=(110:2cm)]0,0) arc (110:120:2cm);

\draw[black,thick,<-] ($(center)+(210:2cm)$)--($(center)+(30:2cm)$);
\coordinate (lu1) at ($(center)+(210:2.3cm)$) {};
\coordinate (lu0) at ($(center)+(30:2.3cm)$) {};
\draw[green,ultra thick,-] ([shift=(200:2cm)]0,0) arc (200:220:2cm);
\draw[red,ultra  thick,-] ([shift=(220:2cm)]0,0) arc (220:230:2cm);
\draw[blue,ultra thick,-] ([shift=(230:2cm)]0,0) arc (230:240:2cm);
\draw[red,ultra  thick,-] ([shift=(190:2cm)]0,0) arc (190:200:2cm);
\draw[blue,ultra  thick,-] ([shift=(180:2cm)]0,0) arc (180:190:2cm);

\draw[black,very thick,<-] ($(center)+(330:2cm)$)--($(center)+(150:2cm)$);
\coordinate (lw1) at ($(center)+(330:2.3cm)$) {};
\coordinate (lw0) at ($(center)+(150:2.3cm)$) {};
\draw[green,ultra  thick,-] ([shift=(320:2cm)]0,0) arc (320:340:2cm);
\draw[red,ultra  thick,-] ([shift=(340:2cm)]0,0) arc (340:350:2cm);
\draw[blue,ultra  thick,-] ([shift=(350:2cm)]0,0) arc (350:360:2cm);
\draw[red,ultra  thick,-] ([shift=(310:2cm)]0,0) arc (310:320:2cm);
\draw[blue,ultra  thick,-] ([shift=(300:2cm)]0,0) arc (300:310:2cm);

\tikzstyle{every node}=[inner sep=1pt]
\begin{tiny}
\node at (lv0) {$s^0$};
\node at (lv1) {$s^1$};
\node at (lu0) {$v^0$};
\node at (lu1) {$v^1$};
\node at (lw0) {$u^0$};
\node at (lw1) {$u^1$};
\end{tiny}

\draw[white] (-2.8,-2.8)--(-2.8,-2.5);
\draw[white] (2.8,2.8)--(2.8,2.5);

\end{tikzpicture}
    \caption{Schematic view of the model $R$ of~$G$}\label{fig:A_counter_N_model}
\end{subfigure}
\,
\begin{subfigure}[b]{0.45\linewidth}
\centering
\begin{tikzpicture}[xscale=0.7,yscale=0.5]

\coordinate (u1) at (-3,2) {};
\coordinate (u2) at (-3.4,1.5) {};
\coordinate (u3) at (-3.4,1.0) {};
\coordinate (u4) at (-3,0.5) {};
\coordinate (u5) at (-2.6,1) {};
\coordinate (u6) at (-2.6,1.5) {};

\coordinate (v1) at (-3,-2) {};
\coordinate (v2) at (-3.4,-1.5) {};
\coordinate (v3) at (-3.4,-1) {};
\coordinate (v4) at (-3,-0.5) {};
\coordinate (v5) at (-2.6,-1) {};
\coordinate (v6) at (-2.6,-1.5) {};

\coordinate (v) at (-1,-1) {};
\coordinate (lv) at (-1,-1.4) {};
\coordinate (u) at (-1,1) {};
\coordinate (lu) at (-1,1.4) {};
\coordinate (w) at (1,0) {};
\coordinate (lw) at (1,0.4) {};

\coordinate (w1) at (3,-1) {};
\coordinate (w2) at (3.4,-0.35) {};
\coordinate (w3) at (3.4,0.35) {};
\coordinate (w4) at (3,1) {};
\coordinate (w5) at (2.6,0.35) {};
\coordinate (w6) at (2.6,-0.35) {};

\coordinate (lV0) at (-3,-3) {};  
\coordinate (lV1) at (-1,-3) {};  
\coordinate (lV2) at (1,-3) {};  
\coordinate (lV3) at (3,-3) {};  
\draw[rounded corners=5, fill=gray!10] (-3.6,-2.5)--(-3.6,2.5) -- (-2.4,2.5) -- (-2.4,-2.5)--cycle;
\draw[rounded corners=5, fill=gray!10] (-1.6,-2.5)--(-1.6,2.5) -- (-0.4,2.5) -- (-0.4,-2.5)--cycle;
\draw[rounded corners=5, fill=gray!10] (0.4,-2.5)--(0.4,2.5) -- (1.6,2.5) -- (1.6,-2.5)--cycle;
\draw[rounded corners=5, fill=gray!10] (2.4,-2.5)--(2.4,2.5) -- (3.6,2.5) -- (3.6,-2.5)--cycle;

\tikzstyle{every node}=[inner sep=1pt]

\path (u1) edge[thick] (u2);
\path (u2) edge[thick] (u3);
\path (u3) edge[thick] (u4);
\path (u4) edge[thick] (u5);
\path (u5) edge[thick] (u6);
\path (u6) edge[thick] (u1);

\path (u) edge[thick] (u1);
\path (u) edge[thick] (u4);

\path (v1) edge[thick] (v2);
\path (v2) edge[thick] (v3);
\path (v3) edge[thick] (v4);
\path (v4) edge[thick] (v5);
\path (v5) edge[thick] (v6);
\path (v6) edge[thick] (v1);

\path (v) edge[thick] (v1);
\path (v) edge[thick] (v4);

\path (w1) edge[thick] (w2);
\path (w2) edge[thick] (w3);
\path (w3) edge[thick] (w4);
\path (w4) edge[thick] (w5);
\path (w5) edge[thick] (w6);
\path (w6) edge[thick] (w1);

\path (w) edge[thick] (w1);
\path (w) edge[thick] (w4);

\path (w) edge[thick] (u);
\path (w) edge[thick] (v);
\path (u) edge[thick] (v);

\tikzstyle{every node}=[circle,minimum size=5pt,inner sep=0pt,draw,fill]
\node[green] at (u1) {};
\node[red] at (u2) {};
\node[blue] at (u3) {};
\node[green] at (u4) {};
\node[blue] at (u5) {};
\node[red] at (u6) {};

\node[green] at (v1) {};
\node[blue] at (v2) {};
\node[red] at (v3) {};
\node[green] at (v4) {};
\node[red] at (v5) {};
\node[blue] at (v6) {};

\node[green] at (w1) {};
\node[blue] at (w2) {};
\node[red] at (w3) {};
\node[green] at (w4) {};
\node[red] at (w5) {};
\node[blue] at (w6) {};

\node[black] at (u) {};
\node[black] at (v) {};
\node[black] at (w) {};

\begin{tiny}
\tikzstyle{every node}=[inner sep=2pt]
\node at (lV0) {$V_0$};
\node at (lV1) {$V_1$};
\node at (lV2) {$V_2$};
\node at (lV3) {$V_3$};

\node at (lu) {$s$};
\node at (lv) {$v$};
\node at (lw) {$u$};
\end{tiny}

\draw[white] (-4,-3.6)--(-4,-2.7);
\draw[white] (4,3.6)--(4,2.7);
\end{tikzpicture}
\caption{Graph $G_c$ and its join decomposition}\label{fig:A_counter_G_c}
  \end{subfigure}

\caption{Counterexample to Claim A.}
\label{fig:A_counter}
\end{figure}
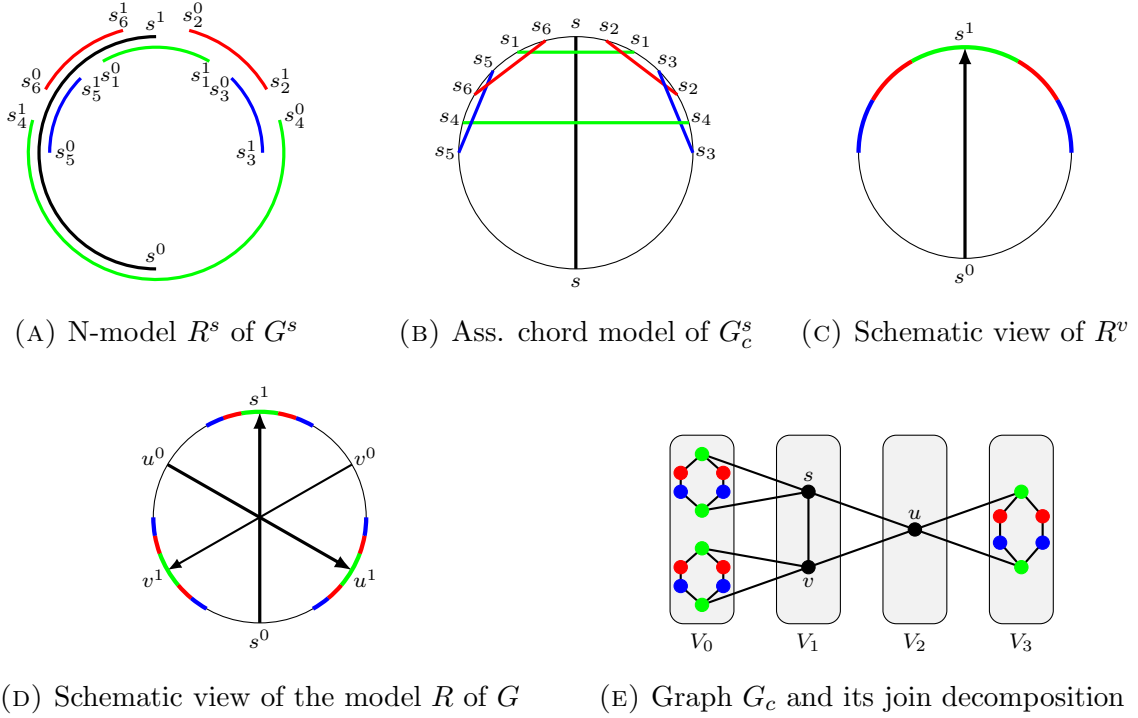

We show that Claim~A is false.
To show a counterexample, we first define an auxiliary circular-arc graph $G^s$: the vertex set 
of $G^s$ is $\{s,s_1,\ldots,s_6\}$, the edges of $G^s$ can be read from the intersection model 
$R^s$ of $G^s$ shown in Figure~\ref{fig:A_counter}(A).
We leave the reader to verify that the model $R^s$ is normalized
(the vertices $s_1,\ldots,s_6$ induce a cycle in $G^s$ and hence every two consecutive vertices of this cycle strictly but not strongly overlap).

Next, we define a circular-arc graph $G$ by taking~$G^s$ and two its isomorphic copies, say~$G^v$ and~$G^u$ over the vertex sets $\{v,v_1,\ldots,v_6\}$ and $\{u,u_1,\ldots,u_6\}$, respectively.
Again, we define the edges of $G$ by providing an intersection model~$R$ of~$G$.
Given the schematic view of the model~$R$ as shown in Figure~\ref{fig:A_counter}(D),
we obtain the intersection model~$R$ for $G$ as follows.
We take the circular-arc model $R^s$ of $G^s$ and we paste its arcs into the circle such that:
\begin{itemize}
 \item the arc $R^s(s)$ has the same endpoints as the oriented chord $s^0s^1$ 
and is on the left side of this chord if we traverse it from $s^0$ to $s^1$,
 \item the endpoints of the remaining arcs of $R^s$ are squeezed such that they all fit 
 the colourful arc adjacent to the head of the chord $s^0s^1$. 
\end{itemize}
We proceed analogously with the models $R^v$ and $R^u$, which represent $G^v$ and $G^u$ in the same way as $R^s$ represents $G^s$.
Again, we leave the reader to verify that $R$ is a normalized model of $G$.
The associated circle graph $G_c$ as well as its join $(V_0,V_1,V_2,V_3)$ are shown in Figure~\ref{fig:A_counter}(E).
Note that:
\begin{itemize}
 \item $G_c$ is s-inseparable as $G_c$ has no non-trivial modules,
 \item vertex $s$ in $V_1$ is adjacent to all other vertices in $V_1 \cup V_2$.
\end{itemize}
That is, $G_c$, $(V_0,V_1,V_2,V_3)$, and $s$ satisfy the assumptions of Claim~A.
The graph $G_c \setminus \{s\}$, however, is not s-inseparable as $G_c \setminus \{s\}$ is disconnected.
So, Claim~A is false.

Then, the proof assumes that $G_c$ has a join $(V_0,V_1,V_2,V_3)$ such that 
no vertex from $V_1 \cup V_2$ is adjacent to all other vertices in $V_1 \cup V_2$. 
Let $H_1$ and $H_2$ be the decomposition of $G_c$ induced by $(V_0,V_1,V_2,V_3)$.
In this setting, the following property of $H_1$ and $H_2$ is claimed. 

\medskip

\noindent \textbf{Claim B [page~425, line -19 to -16]:} \emph{Therefore, assume no such vertex $s$ exists. 
Then it is easy to verify that the corresponding two component graphs, $H_1$ and $H_2$ of $G_c$
(the existence of the substitution in either $H_1$ or $H_2$ would imply the existence of one in $G_c$), are $s$-inseparable.}

\medskip

\begin{figure}[h]
  \centering \small
  \begin{subfigure}[b]{0.49\linewidth}
    \centering
\begin{tikzpicture}[scale=0.8,>=latex]

\coordinate (center) at (0.0,0.0) {};

\draw[red,very thick,-] ([shift=(200:2cm)]0,0) arc (200:340:2cm);
\coordinate (lu11) at ($(center)+(200:1.7cm)$) {};
\coordinate (lu01) at ($(center)+(340:1.7cm)$) {};
\draw[red,very thick,-] ([shift=(20:2cm)]0,0) arc (20:160:2cm);
\coordinate (lu12) at ($(center)+(20:1.7cm)$) {};
\coordinate (lu02) at ($(center)+(160:1.7cm)$) {};
\draw[green,very thick,-] ([shift=(-40:2.2cm)]0,0) arc (-40:40:2.2cm);
\coordinate (lu03) at ($(center)+(40:2.5cm)$) {};
\coordinate (lu13) at ($(center)+(-40:2.5cm)$) {};
\draw[blue,very thick,-] ([shift=(-30:2.3cm)]0,0) arc (-30:10:2.3cm);
\coordinate (lu04) at ($(center)+(10:2.7cm)$) {};
\coordinate (lu14) at ($(center)+(-30:2.6cm)$) {};
\draw[brown,very thick,-] ([shift=(-10:2.4cm)]0,0) arc (-10:30:2.4cm);
\coordinate (lu15) at ($(center)+(-10:2.7cm)$) {};
\coordinate (lu05) at ($(center)+(30:2.7cm)$) {};

\draw[black,very thick,-] ([shift=(110:2.1cm)]0,0) arc (110:250:2.1cm);
\coordinate (lv11) at ($(center)+(110:1.7cm)$) {};
\coordinate (lv01) at ($(center)+(250:2.4cm)$) {};
\draw[black,very thick,-] ([shift=(290:2.1cm)]0,0) arc (290:430:2.1cm);
\coordinate (lv12) at ($(center)+(290:2.4cm)$) {};
\coordinate (lv02) at ($(center)+(430:1.7cm)$) {};
\draw[green,very thick,-] ([shift=(50:2.2cm)]0,0) arc (50:130:2.2cm);
\coordinate (lv13) at ($(center)+(50:2.5cm)$) {};
\coordinate (lv03) at ($(center)+(130:2.5cm)$) {};
\draw[blue,very thick,-] ([shift=(60:2.3cm)]0,0) arc (60:100:2.3cm);
\coordinate (lv14) at ($(center)+(60:2.6cm)$) {};
\coordinate (lv04) at ($(center)+(100:2.7cm)$) {};
\draw[brown,very thick,-] ([shift=(80:2.4cm)]0,0) arc (80:120:2.4cm);
\coordinate (lv15) at ($(center)+(80:2.7cm)$) {};
\coordinate (lv05) at ($(center)+(120:2.7cm)$) {};

\tikzstyle{every node}=[inner sep=1pt]
\begin{tiny}
\node at (lu01) {$u^0_1$};
\node at (lu11) {$u^1_1$};

\node at (lu02) {$u^0_2$};
\node at (lu12) {$u^1_2$};

\node at (lu03) {$u^0_3$};
\node at (lu13) {$u^1_3$};

\node at (lu04) {$u^0_4$};
\node at (lu14) {$u^1_4$};

\node at (lu05) {$u^0_5$};
\node at (lu15) {$u^1_5$};

\node at (lv01) {$v^0_1$};
\node at (lv11) {$v^1_1$};

\node at (lv02) {$v^0_2$};
\node at (lv12) {$v^1_2$};

\node at (lv03) {$v^0_3$};
\node at (lv13) {$v^1_3$};

\node at (lv04) {$v^0_4$};
\node at (lv14) {$v^1_4$};

\node at (lv05) {$v^0_5$};
\node at (lv15) {$v^1_5$};
\end{tiny}

\draw[white] (-3,-3)--(-3,-2.7);
\draw[white] (3,3)--(3,2.7);

\end{tikzpicture}
    \caption{Normalized model of~$G$}\label{fig:B_counter_normalized_model}
  \end{subfigure}
  \,
  \begin{subfigure}[b]{0.49\linewidth}
    \centering
\begin{tikzpicture}[scale=0.8,>=latex]
\coordinate (center) at (0.0,0.0) {};
\draw (center) circle (2.2cm);

\draw[red,very thick,-] ($(center)+(200:2.2cm)$)--($(center)+(340:2.2cm)$);
\coordinate (lu11) at ($(center)+(200:2.45cm)$) {};
\coordinate (lu01) at ($(center)+(340:2.45cm)$) {};
\draw[red,very thick,-] ($(center)+(20:2.2cm)$)--($(center)+(160:2.2cm)$);
\coordinate (lu12) at ($(center)+(20:2.45cm)$) {};
\coordinate (lu02) at ($(center)+(160:2.45cm)$) {};
\draw[green,very thick,-] ($(center)+(-40:2.2cm)$)--($(center)+(40:2.2cm)$);
\coordinate (lu13) at ($(center)+(-40:2.45cm)$) {};
\coordinate (lu03) at ($(center)+(40:2.45cm)$) {};
\draw[blue,very thick,-] ($(center)+(-30:2.2cm)$)--($(center)+(10:2.2cm)$);
\coordinate (lu14) at ($(center)+(10:2.45cm)$) {};
\coordinate (lu04) at ($(center)+(-30:2.45cm)$) {};
\draw[brown,very thick,-] ($(center)+(-10:2.2cm)$)--($(center)+(30:2.2cm)$);
\coordinate (lu15) at ($(center)+(-10:2.45cm)$) {};
\coordinate (lu05) at ($(center)+(30:2.45cm)$) {};

\draw[black,very thick,-] ($(center)+(110:2.2cm)$)--($(center)+(250:2.2cm)$);
\coordinate (lv12) at ($(center)+(110:2.45cm)$) {};
\coordinate (lv02) at ($(center)+(250:2.45cm)$) {};
\draw[black,very thick,-] ($(center)+(290:2.2cm)$)--($(center)+(430:2.2cm)$);
\coordinate (lv11) at ($(center)+(290:2.45cm)$) {};
\coordinate (lv01) at ($(center)+(430:2.45cm)$) {};
\draw[green,very thick,-] ($(center)+(50:2.2cm)$)--($(center)+(130:2.2cm)$);
\coordinate (lv13) at ($(center)+(50:2.45cm)$) {};
\coordinate (lv03) at ($(center)+(130:2.45cm)$) {};
\draw[blue,very thick,-] ($(center)+(60:2.2cm)$)--($(center)+(100:2.2cm)$);
\coordinate (lv14) at ($(center)+(60:2.45cm)$) {};
\coordinate (lv04) at ($(center)+(100:2.45cm)$) {};
\draw[brown,very thick,-] ($(center)+(80:2.2cm)$)--($(center)+(120:2.2cm)$);
\coordinate (lv15) at ($(center)+(80:2.45cm)$) {};
\coordinate (lv05) at ($(center)+(120:2.45cm)$) {};

\tikzstyle{every node}=[inner sep=1pt]
\begin{tiny}
\node at (lu01) {$u_1$};
\node at (lu11) {$u_1$};

\node at (lu02) {$u_2$};
\node at (lu12) {$u_2$};

\node at (lu03) {$u_3$};
\node at (lu13) {$u_3$};

\node at (lu04) {$u_4$};
\node at (lu14) {$u_4$};

\node at (lu05) {$u_5$};
\node at (lu15) {$u_5$};

\node at (lv01) {$v_1$};
\node at (lv11) {$v_1$};

\node at (lv02) {$v_2$};
\node at (lv12) {$v_2$};

\node at (lv03) {$v_3$};
\node at (lv13) {$v_3$};

\node at (lv04) {$v_4$};
\node at (lv14) {$v_4$};

\node at (lv05) {$v_5$};
\node at (lv15) {$v_5$};
\end{tiny}

\draw[white] (-3,-3)--(-3,-2.7);
\draw[white] (3,3)--(3,2.7);

\end{tikzpicture}
    \caption{Associated chord model $D$ of $G_c$}\label{fig:ca-graph}
  \end{subfigure}
\vspace{0.2cm}

\begin{subfigure}[b]{0.45\linewidth}
\centering
\begin{tikzpicture}[xscale=0.7,yscale=0.5]

\coordinate (lu3) at (-1.8,0.5) {};
\coordinate (u3) at (-1.8,0) {};
\coordinate (u4) at (-2.2,1.2) {};
\coordinate (lu4) at (-2.2,1.7) {};
\coordinate (u5) at (-2.2,-1.2) {};
\coordinate (lu5) at (-2.2,-1.7) {};

\coordinate (lu1) at (0,1.7) {};
\coordinate (u1) at (0,1.2) {};
\coordinate (u2) at (0,-1.2) {};
\coordinate (lu2) at (0,-1.7) {};

\coordinate (lv1) at (2,1.7) {};
\coordinate (v1) at (2,1.2) {};
\coordinate (v2) at (2,-1.2) {};
\coordinate (lv2) at (2,-1.7) {};

\coordinate (lv3) at (3.8,0.5) {};
\coordinate (v3) at (3.8,0) {};
\coordinate (v4) at (4.2,1.2) {};
\coordinate (lv4) at (4.2,1.7) {};
\coordinate (v5) at (4.2,-1.2) {};
\coordinate (lv5) at (4.2,-1.7) {};

\coordinate (lV0) at (-2,-3) {};  
\coordinate (lV1) at (0,-3) {};  
\coordinate (lV2) at (2,-3) {};  
\coordinate (lV3) at (4,-3) {};  
\draw[rounded corners=5, fill=gray!10] (-2.6,-2.5)--(-2.6,2.5) -- (-1.4,2.5) -- (-1.4,-2.5)--cycle;
\draw[rounded corners=5, fill=gray!10] (-0.6,-2.5)--(-0.6,2.5) -- (0.6,2.5) -- (0.6,-2.5)--cycle;
\draw[rounded corners=5, fill=gray!10] (1.4,-2.5)--(1.4,2.5) -- (2.6,2.5) -- (2.6,-2.5)--cycle;
\draw[rounded corners=5, fill=gray!10] (3.4,-2.5)--(3.4,2.5) -- (4.6,2.5) -- (4.6,-2.5)--cycle;

\tikzstyle{every node}=[circle,minimum size=5pt,inner sep=0pt,draw,fill]
\node at (u1) {};
\node at (u2) {};
\node at (u3) {};
\node at (u4) {};
\node at (u5) {};
\node at (v1) {};
\node at (v2) {};
\node at (v3) {};
\node at (v4) {};
\node at (v5) {};

\tikzstyle{every node}=[inner sep=1pt]

\path (u1) edge[thick] (u4);
\path (u1) edge[thick] (u3);
\path (u2) edge[thick] (u5);
\path (u2) edge[thick] (u3);
\path (u4) edge[thick] (u5);

\path (u1) edge[thick] (v1);
\path (u1) edge[thick] (v2);
\path (u2) edge[thick] (v1);
\path (u2) edge[thick] (v2);

\path (v1) edge[thick] (v4);
\path (v1) edge[thick] (v3);
\path (v2) edge[thick] (v5);
\path (v2) edge[thick] (v3);
\path (v4) edge[thick] (v5);

\begin{tiny}
\tikzstyle{every node}=[inner sep=2pt]
\node at (lV0) {$V_0$};
\node at (lV1) {$V_1$};
\node at (lV2) {$V_2$};
\node at (lV3) {$V_3$};

\node at (lu1) {$u_1$};
\node at (lu2) {$u_2$};
\node at (lu3) {$u_3$};
\node at (lu4) {$u_4$};
\node at (lu5) {$u_5$};

\node at (lv1) {$v_1$};
\node at (lv2) {$v_2$};
\node at (lv3) {$v_3$};
\node at (lv4) {$v_4$};
\node at (lv5) {$v_5$};

\end{tiny}
\end{tikzpicture}
\caption{Graph $G_c$ and its join decomposition}\label{fig:ca-graph}
  \end{subfigure}
\,
\begin{subfigure}[b]{0.48\linewidth}
\centering
\begin{tikzpicture}[xscale=0.7,yscale=0.5]

\coordinate (lu3) at (-1.8,0.5) {};
\coordinate (u3) at (-1.8,0) {};
\coordinate (u4) at (-2.2,1.2) {};
\coordinate (lu4) at (-2.2,1.7) {};
\coordinate (u5) at (-2.2,-1.2) {};
\coordinate (lu5) at (-2.2,-1.7) {};

\coordinate (lu1) at (0,1.7) {};
\coordinate (u1) at (0,1.2) {};
\coordinate (u2) at (0,-1.2) {};
\coordinate (lu2) at (0,-1.7) {};

\coordinate (lv) at (2,0.6) {};
\coordinate (v) at (2,0) {};

\coordinate (lV0) at (-2,-3) {};  
\coordinate (lV1) at (0,-3) {};  
\draw[rounded corners=5, fill=gray!10] (-2.6,-2.5)--(-2.6,2.5) -- (-1.4,2.5) -- (-1.4,-2.5)--cycle;
\draw[rounded corners=5, fill=gray!10] (-0.6,-2.5)--(-0.6,2.5) -- (0.6,2.5) -- (0.6,-2.5)--cycle;

\tikzstyle{every node}=[circle,minimum size=5pt,inner sep=0pt,draw,fill]
\node at (u1) {};
\node at (u2) {};
\node at (u3) {};
\node at (u4) {};
\node at (u5) {};
\node at (v) {};

\tikzstyle{every node}=[inner sep=1pt]

\path (u1) edge[thick] (u4);
\path (u1) edge[thick] (u3);
\path (u2) edge[thick] (u5);
\path (u2) edge[thick] (u3);
\path (u4) edge[thick] (u5);

\path (u1) edge[thick] (v);
\path (u2) edge[thick] (v);

\begin{tiny}
\tikzstyle{every node}=[inner sep=2pt]
\node at (lV0) {$V_0$};
\node at (lV1) {$V_1$};

\node at (lu1) {$u_1$};
\node at (lu2) {$u_2$};
\node at (lu3) {$u_3$};
\node at (lu4) {$u_4$};
\node at (lu5) {$u_5$};

\node at (lv) {$v$};

\end{tiny}
\end{tikzpicture}
\begin{tikzpicture}[xscale=0.7,yscale=0.5]

\coordinate (lu) at (0,0.6) {};
\coordinate (u) at (0,0) {};

\coordinate (lv1) at (2,1.7) {};
\coordinate (v1) at (2,1.2) {};
\coordinate (v2) at (2,-1.2) {};
\coordinate (lv2) at (2,-1.7) {};

\coordinate (lv3) at (3.8,0.5) {};
\coordinate (v3) at (3.8,0) {};
\coordinate (v4) at (4.2,1.2) {};
\coordinate (lv4) at (4.2,1.7) {};
\coordinate (v5) at (4.2,-1.2) {};
\coordinate (lv5) at (4.2,-1.7) {};

\coordinate (lV2) at (2,-3) {};  
\coordinate (lV3) at (4,-3) {};  
\draw[rounded corners=5, fill=gray!10] (1.4,-2.5)--(1.4,2.5) -- (2.6,2.5) -- (2.6,-2.5)--cycle;
\draw[rounded corners=5, fill=gray!10] (3.4,-2.5)--(3.4,2.5) -- (4.6,2.5) -- (4.6,-2.5)--cycle;

\tikzstyle{every node}=[circle,minimum size=5pt,inner sep=0pt,draw,fill]
\node at (u) {};
\node at (v1) {};
\node at (v2) {};
\node at (v3) {};
\node at (v4) {};
\node at (v5) {};

\tikzstyle{every node}=[inner sep=1pt]

\path (u) edge[thick] (v1);
\path (u) edge[thick] (v2);

\path (v1) edge[thick] (v4);
\path (v1) edge[thick] (v3);
\path (v2) edge[thick] (v5);
\path (v2) edge[thick] (v3);
\path (v4) edge[thick] (v5);

\begin{tiny}
\tikzstyle{every node}=[inner sep=2pt]
\node at (lV2) {$V_2$};
\node at (lV3) {$V_3$};

\node at (lu) {$u$};

\node at (lv1) {$v_1$};
\node at (lv2) {$v_2$};
\node at (lv3) {$v_3$};
\node at (lv4) {$v_4$};
\node at (lv5) {$v_5$};

\end{tiny}
\end{tikzpicture}

\caption{Decomposition of $G_c$ into $H_1$ and $H_2$}\label{fig:ca-graph}
\end{subfigure}

\caption{Counterexample to Claim B.}
\label{fig:B_counter}

\end{figure}
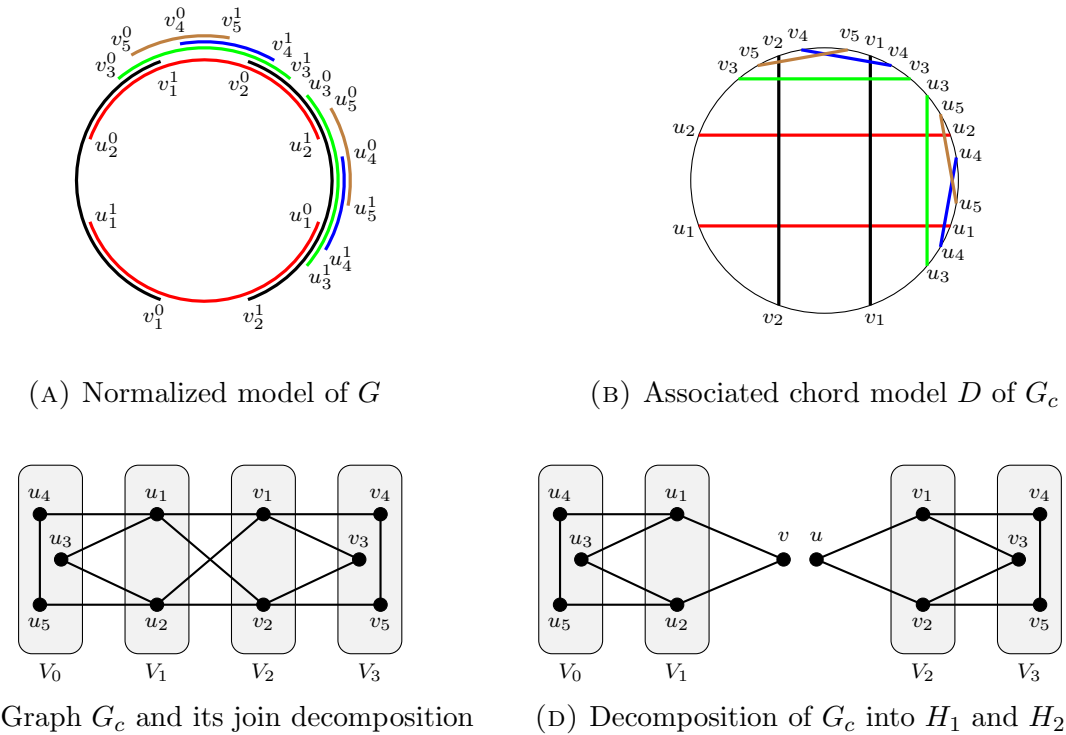

Consider a circular-arc graph $G$ whose normalized model $R$ is shown in Figure~\ref{fig:B_counter}(A).
The associated circle graph $G_c$ and its join $(V_0,V_1,V_2,V_3)$ are shown in Figure~\ref{fig:B_counter}(C),
the chord model $D$ associated with $R$ is shown in Figure~\ref{fig:B_counter}(B),
and the decomposition $H_1$ and $H_2$ of $G_c$ induced by the join $(V_0,V_1,V_2,V_3)$ is shown in Figure~\ref{fig:B_counter}(D).
Note the following properties of $G_c$, $H_1$, and $H_2$:
\begin{itemize}
 \item $G_c$ is s-inseparable as $G_c$ has no non-trivial modules and there is no vertex $s$ in $V_1 \cup V_2$ adjacent to all other vertices in $V_1 \cup V_2$,
 \item $H_1$ contains a non-trivial module $\{u_3,v\}$ and $H_2$ contains a non-trivial module $\{u,v_3\}$.
\end{itemize}
In particular, Claim B is also false.

The ``proof'' of Theorem 5.4 is concluded as follows.
Since $H_1$ and $H_2$ are s-inseparable, they have unique chord models by induction.
Eventually, there is a unique way to compose those models to get a conformal model of $G_c$.

\subsection{Consistent modules}
\label{subsec:consistent_modules}

The precise definition of the consistent modules (they are also called as T-submodules in~\cite{Hsu95}) 
of $G_c$ is given in~Section 6.1.
Since it is very technical, we list only their properties as proven by Hsu.

\smallskip
\noindent \textbf{Lemma 6.5 [page 428, -2 -- page 429, line 2]}
\emph{Let $M$ be a T-submodule that gives rise to a permutation graph. 
Let $D$ be any conformal model for $G_c$. 
Then the endpoints of chords of $M$ can be partitioned into two
sets $A_1$, $A_2$ such that each chord of $M$ has one endpoint in $A_1$ and the other in $A_2$ and those
endpoints in $A_1$ (resp., $A_2$) are consecutive in $D$.}

\smallskip
\noindent \textbf{Theorem 6.6 [page 429, lines -15 to -11]} 
\emph{Consider a neighborhood module $V(G_c)$. 
Let $V' = \{v_1,\ldots, v_r\}$ be a set of representatives from the maximal consistent submodules $M^1,\ldots,M^p$\footnote{The statement of Theorem 6.6 contains a typo -- we should have $r=p$.
} in the consistent partition of $V(G_c)$. 
Then there is a unique conformal model for $V'$. 
Furthermore, this model is independent of the $v_i$ selected 
(we shall refer to the subgraph induced on $V'$ as the
representative graph for $V(G)$).}

In fact, the above two lemmas would assert Property~\ref{prop:H3} of the consistent modules of~$G_c$.
Note that, if $G_c$ is a permutation graph (as it is for the case when $G$ is co-bipartite), then one could take the children of $V(G_c)$ as the consistent modules of~$G_c$ (all modules in the modular decomposition tree of $G_c$ are ``consistent'' in all permutation models of~$G_c$).
Hence, to determine the consistent modules of $G_c$ Hsu focuses his attention on the children of $V(G_c)$ in 
the modular decomposition tree of $G_c$.
Hsu rightly notices that they might be no longer consistent when $G_c$ is not a permutation graph.
In fact, assuming that $M_1,\ldots,M_k$ are the children of $V(G_c)$ in the modular decomposition tree of $G_c$,
Hsu ``proves'' the following lemma:

\smallskip
\noindent \textbf{Lemma~6.3 [Page 428, lines 11--12]} \emph{If a maximal submodule $M_i$ of $V(G_c)$ is not consistent, then it is a series
module.}

However, the above lemma is also false, as parallel children of $V(G)$ might also be not consistent.
This enables us to construct the graph claimed by Counterexample~\ref{count:main_counter}.

\smallskip
\noindent \textbf{Counterexample~\ref{count:main_counter}}.
Consider a circular-arc graph $G$ shown in Figure~\ref{fig:ca_graph_transformation}.(A). 
Its normalized model~$R$ is shown in Figure~\ref{fig:ca_graph_transformation}.(B), an associated chord model~$D$ in Figure~\ref{fig:ca_graph_transformation}.(C), and the circle graph~$G_c$ associated to~$G$ and its partition into maximal non-trivial modules $M_1,\ldots,M_4$ in Figure~\ref{fig:ca_graph_transformation}.(D).
Note that $V(G)$ is the neighbourhood module in the modular decomposition tree of~$G_c$, and each $M_i$ is a parallel child of $V(G)$.
Note that the modules $M_1$ and $M_4$ are not consistent in $D$.
One can also check that the model $R$ and its reflection are the only two normalized models of $G$.
Since both of them do not follow the description given by Hsu (it is claimed in~\cite{Hsu95} that only series children of $V(G)$ might be not consistent),
the graph~$G$ proves the statement of Counterexample~\ref{count:main_counter}.

The graph $G$ also shows that, assuming Hsu's partition into ``consistent modules'', Lemma~6.5 and Theorem~6.6 from~\cite{Hsu95} are also false.

\section{$V(G_c)$ is a parallel module}
\label{sec:parallel_case}

In the case where $V(G_c)$ is a parallel module, 
the description given by Hsu is incomplete (and also not correct).
In particular, Hsu constructs so-called ``consistent module tree'' for each component $C$
of $G_c$ basing on the consistent modules (T-submodules) of the component $C$.
As is written in \cite{Hsu95} (see page 435, lines 27 to 29):
\emph{We shall first construct a consistent decomposition tree~$T$ for the subgraph~$G_c[C]$ 
and then insert the vertices of $K$ into conformal models of appropriate T-submodules. 
For any T-submodule $M$ in $C$, let .....} and then a description of the tree $T$ for the component $C$ follows.

Each component~$C$ of $G_c$ is either a neighborhood or a series module in $G_c$.
In the case when $C$ is a neighborhood component in $G_c$, the consistent modules (T-modules) of~$C$, as we have seen in the previous section, are not determined correctly by Hsu.
Nowhere in~\cite{Hsu95} we found how Hsu defines consistent modules (T-modules) for series components of~$G_c$.

\bibliographystyle{plain}
\bibliography{../lit_short}

\end{document}